%% file: main.tex
\documentclass[aps,prd,twocolumn,10pt,nofootinbib,superscriptaddress,preprintnumbers,floatfix]{revtex4-1}
\pdfoutput=1 
\usepackage[utf8]{inputenc} 
\usepackage{amsmath} 
\usepackage{amsfonts}
\usepackage{amsthm}
\usepackage{amssymb}
\usepackage{graphicx} 
\usepackage{xcolor} 
\usepackage[normalem]{ulem} 
\usepackage{slashed} 
\usepackage{physics} 
\usepackage{mathrsfs} 
\usepackage{hyperref} 
\usepackage{comment}

\graphicspath{{./}{./figs/}}

\newcommand{\rmii}[1]{{\mbox{\tiny\rm{#1}}}}

\newcommand{\ddx}{\dd^3\mathbf{x}}
\newcommand{\ddp}{\dd^3\mathbf{p}}
\newcommand{\feq}{f_{\text{eq},a}}
\newcommand{\VCB}{V_\text{CB}}

\newcommand{\HCB}{H^\text{CB}_\text{const}}
\newcommand{\hquad}{H^\text{CB}_\text{quad}}
\newcommand{\hfluct}{H^\text{CB}_\text{fluct}}

\newcommand{\Hmeta}{H^\text{meta}_\text{const}}
\newcommand{\hmetaquad}{H^\text{meta}_\text{quad}}
\newcommand{\hmetafluct}{H^\text{meta}_\text{fluct}}

\newcommand{\DHCB}{\Delta\!\;\! H_\text{CB}}

\newcommand{\dfa}{\delta\!f_a}

\newcommand{\udp}{U_{\delta\phi}}

\newcommand{\ldp}{\overline{\delta\phi}}
\newcommand{\lp}{\overline{\pi}}
\newcommand{\ldf}{\overline{\dfa}}
\newcommand{\funcdv}[1]{\frac{\delta}{\delta#1}}

\newcommand{\cb}{\phi_{\rmii{CB}}}
\newcommand{\vnucl}{V_{\text{nucl}}}

\usepackage{color} 
\definecolor{CORRcolour}{rgb}{1,0.435,0}

\definecolor{FinnishBlue}{rgb}{0,0.435,1}

\definecolor{HeavyColour}{rgb}{0.2,0.77,0}

\newcommand{\Helsinki}{\affiliation{
    Department of Physics and Helsinki Institute of Physics,
    P.O.\ Box 64,
    FI-00014 University of Helsinki,
    Finland
}}

\newcommand{\Nottingham}{\affiliation{
    School of Physics and Astronomy, University of Nottingham, Nottingham NG7 2RD, United Kingdom
}}

\begin{document}

\title{Real-Time Nucleation and Off-Equilibrium Effects\\in High-Temperature Quantum Field Theories}
\date{June 26, 2025}

\author{Joonas Hirvonen}
\email{joonas.hirvonen@nottingham.ac.uk}
\Helsinki
\Nottingham

\begin{abstract}
We study real-time nucleation in perturbative high-temperature quantum field theories. Specifically, we incorporate the evolution of thermally fluctuating plasma driven out of equilibrium by nucleation. This plasma forms the thermal bath for the nucleating bubbles, and its off-equilibrium dynamics backreact on the bubbles, modifying the nucleation rate. Utilizing kinetic Boltzmann descriptions for the plasma particles, we derive that the nucleation rate in high-temperature quantum field theories is described by Langer's rate formula with effects arising from the off-equilibrium plasma, unaccounted for in Linde's thermal rate. Importantly, we establish a connection to equilibrium computations of nucleation rates, particularly with the effective field theory approach, showing that they correctly capture the leading parts of the logarithm of the rate. We also show that the dominant modifications to the nucleation rate from a light bosonic quantum field arise from its corresponding long-wavelength classical modes, rather than the associated hard particles. This work provides a foundation for further studies of off-equilibrium effects in nucleation rates for realistic Standard Model extensions.
\end{abstract}

\preprint{HIP-2024-6/TH}
\maketitle

\tableofcontents

\section{Introduction}

First-order phase transitions occurring during the cosmological evolution of our Universe give a tantalizing possibility for signals from physics beyond the Standard Model. These transitions may have produced a stochastic gravitational wave background~\cite{Witten:1984rs,Hogan:1986dsh,Caprini:2018mtu,Caprini:2019egz,Hindmarsh:2020hop,Athron:2023xlk} that could be observable with gravitational-wave detectors~\cite{Arzoumanian:2020vkk,Audley:2017drz,Kawamura:2011zz,Harry:2006fi,Guo:2018npi}. Additionally, a first-order electroweak phase transition could have produced the baryon asymmetry of the Universe through baryogenesis~\cite{Kuzmin:1985mm,Shaposhnikov:1986jp,Shaposhnikov:1987tw,Morrissey:2012db}.

First-order phase transitions occur via the nucleation of bubbles of a new, stable phase. Hence, the rate of nucleation determines central properties of the transitions, such as their durations and transition temperatures~\cite{Enqvist:1991xw,Ellis:2018mja}. Both of these quantities are needed for predicting gravitational-wave spectra resulting from the transitions~\cite{Hindmarsh:2017gnf}. In particular, inaccuracies in nucleation rates are one of the current bottlenecks in accurately predicting the spectrum from a transition~\cite{Gould:2021oba}. Hence, a precise understanding of nucleation rates is crucial if we want to connect possible beyond-the-Standard-Model physics with future observations.

The history of studying nucleation rates in field theories began in the late 60s with Langer developing classical nucleation theory~\cite{Langer:1967ax,Langer:1969bc,langer1974metastable}. A decade later, Coleman and Callan created a formalism for vacuum decay in quantum field theories (QFTs)~\cite{Coleman:1977py,Callan:1977pt}, which was subsequently generalized to finite-temperature tunneling by Linde~\cite{Linde:1981zj}. Also, a lattice framework was formulated for numerical nucleation-rate studies in high-temperature QFTs~\cite{Moore:2000jw,Moore:2001vf},%
\footnote{The lattice formulation~\cite{Moore:2000mx} can take into account very similar effects as studied in this article. There, hard gauge-boson particles induce Markovian noise and dissipation onto the evolution of the corresponding classical gauge fields in non-Abelian gauge theories~\cite{Bodeker:1998hm}. Here, the hard particles affect the nucleating fields directly, and the description is not Markovian in general.}
based on classical nucleation theory.
(See Refs.~\cite{Gould:2022ran,Gould:2024chm} for more recent lattice results.)

Recently, there has been progress in analytical understanding and semianalytical computations of high-temperature nucleation rates based on classical nucleation theory. The effective field theory (EFT) approach~\cite{Gould:2021ccf} makes an explicit connection to the classical nucleation theory, albeit constrained to an equilibrium part of the nucleation rate. The rate has been confirmed to be gauge invariant~\cite{Hirvonen:2021zej,Lofgren:2021ogg}, the convergence has been studied~\cite{Ekstedt:2022ceo}, and the ensuing fluctuation determinants can be computed with a Python package~\cite{Ekstedt:2023sqc}. Also, Langer's classical rate has been extended to encompass all orders in perturbation theory~\cite{Ekstedt:2022tqk}, and Refs.~\cite{Pirvu:2021roq,Batini:2023zpi,Pirvu:2023plk,Pirvu:2024ova,Pirvu:2024nbe} have studied real-time nucleation numerically using a classical field theory.

Still, a direct real-time computation in high-temperature QFTs is missing from the literature, which we shall provide here. We study the thermal system in real time, taking into account the effects of off-equilibrium plasma on nucleating bubbles with kinetic, Boltzmann-equation descriptions, which can be derived as the appropriate effective descriptions for nonequilibrium QFTs~\cite{Calzetta:1986cq,Jeon:1994if,Arnold:1997gh,Blaizot:2001nr}.

The form of the final result for high-temperature QFTs coincides with Langer's rate formula. Consequently, the present computation can be regarded as a proof of the range of applicability of Langer's rate formula -- originally derived for classical systems with Markovian noise -- extending to high-temperature QFTs, where external fluctuations and dissipation arise from thermal particles.

Importantly, our analysis shows that the statistical, equilibrium part of the nucleation rates remains unchanged in the real-time computation. This validates previous equilibrium studies of nucleation, and makes a concrete link in particular to the EFT approach.

The thermal plasma particles only modify a parameter in the final rate formula, the exponential growth rate of the critical bubble, which is absent from Linde's thermal rate.

In Sec.~\ref{sec:ModelAndVlasov}, we present a QFT model for concreteness and discuss its effective thermal real-time description.
In Sec.~\ref{sec:hamiltonian}, we discover the Hamiltonian for the description and the corresponding equilibrium distribution.
In Sec.~\ref{sec:configurations}, we discuss two important configurations related to nucleation, the critical bubble and the exponentially growing configuration on the critical bubble.
The exponential growth is further investigated in Sec.~\ref{sec:exponentialGrothUltrarelPlasma} for the case of purely ultrarelativistic plasma.
Then in Secs.~\ref{sec:nucleationDistribution} and \ref{sec:offEquilDir}, we follow the seminal works of Kramers~\cite{Kramers:1940zz} and Langer~\cite{Langer:1969bc}, and find the phase-space probability distribution that describes nucleation.
In Sec.~\ref{sec:nucleationRate}, we finally compute the nucleation rate based on the nucleating probability distribution.
We then conclude in Sec.~\ref{sec:conclusionanddiscussion}, and discuss the generality of the results.

\section{Example Model and its Effective Real-Time Description}\label{sec:ModelAndVlasov}

For concreteness, we first introduce a simple model for the computation. The results are more general than the particular model, which is hopefully clear from the presentation of the analysis. This will be discussed further in Sec.~\ref{sec:conclusionanddiscussion}. We present the real-time, high-temperature effective description corresponding to the QFT model. It contains a long-range classical bosonic field, which undergoes nucleation, and thermal particles, which provide thermal kicks and dissipation for the field.

The simple model has a real scalar, $\Phi$, and a Dirac fermion $\Psi$, and is described by the Lagrangian
\begin{align}
    \mathscr{L}&=\mathscr{L}_\Phi+\mathscr{L}_\Psi-y\Phi\Bar{\Psi}\Psi,\label{eq:initialLagrangian}\\
    \mathscr{L}_\Phi&=-\frac{1}{2}\Phi\Box\Phi-\frac{m_\text{b}^2}{2}\Phi^2-\frac{g}{3!}\Phi^3-\frac{\lambda}{4!}\Phi^4,\\
    \mathscr{L}_\Psi&=\Bar{\Psi}(i\slashed{\partial}-m_\text{f})\Psi.
\end{align}
During nucleation, the background field changes significantly, and hence
the background contributions to the masses are important,
\begin{align}
    m_\Phi^2&=V''(\Phi),\label{eq:scalarMass}\\
    m_\Psi&=m_\text{f}+y\Phi.\label{eq:fermionMass}
\end{align}
We lay down power-counting rules containing no hierarchies in couplings or masses,
\begin{gather}
    \lambda\sim g^2/m_\Phi^2\sim y^2\ll 1,\label{eq:perturbativity}\\
    m_\Phi^2\sim m_\Psi^2\sim \lambda T^2\ll T^2.\label{eq:noMassHierarchies}
\end{gather}
Equation~\eqref{eq:perturbativity} will naturally lead to a high-temperature hierarchy at the phase transition~\cite{Gould:2021ccf,Hirvonen:2022jba}, shown in the latter equation.

A thermal phase transition and the ensuing nucleation are not manifestly present in the Lagrangian. The nucleation occurs on a length scale much longer than the characteristic thermal fluctuations, $L_\text{nucl}\sim m_\Phi^{-1}\gg T^{-1}$~\cite{Gould:2021ccf}. Consequently, the nucleating degrees of freedom are the long-wavelength modes of the scalar field, $\phi$, with spatial Fourier momenta of $\mathbf{k}\sim m_\Phi^{-1}\ll T^{-1}$, and we should therefore construct an effective description for $\phi$ from the underlying QFT.

The effective description for $\phi$ simplifies due to the field behaving nearly classically~\cite{Mueller:2002gd,Greiner:1996dx,Aarts:1996qi,Aarts:1997kp,Bodeker:1996wb,Ghiglieri:2020dpq}. This is a consequence of large occupation numbers, $n_B=(e^{E/T}-1)^{-1}\approx T/E \gg 1$, which follow from the high-temperature scale hierarchy. Hence, the quantum effects in the time evolution are subleading. In general, each light bosonic quantum field would have a corresponding long-wavelength classical field.

In Ref.~\cite{Gould:2021ccf}, it was argued that one should construct the equilibrium EFT for the length scale of nucleation, $L_\text{nucl}\sim m_\phi^{-1}$. Here, this EFT construction corresponds to the \textit{high-temperature dimensional reduction} in the imaginary-time formalism~\cite{Farakos:1994kx,Kajantie:1995dw,Braaten:1995cm,Braaten:1995jr,Ekstedt:2022bff,Hirvonen:2022jba}, and to the potential
\begin{align}
    \vnucl&=s_\text{nucl}\phi+\frac{1}{2}m_\text{nucl}^2\phi^2+\frac{g}{3!}\phi^3+\frac{\lambda}{4!}\phi^4,\label{eq:nucleationPotential}\\
    s_\text{nucl}&=(g+4ym_\text{f})T^2/24,\\
    m_\text{nucl}^2&=m_\text{b}^2+(\lambda+4y^2)T^2/24.
\end{align}
Notice that the thermal dependency of the potential initiates the thermal phase transition. We will confirm in Sec.~\ref{sec:hamiltonian} that the equilibrium potential, $\vnucl$, is correct for the real-time analysis.

As we are studying the real-time evolution of the system, we mustn't assume that the thermal particles remain in strict equilibrium. This is reflected in the full equation of motion (EoM),
\begin{equation}\label{eq:startingPointForFields}
    \Box \phi + \vnucl'(\phi)=-\sum_{a}\dv{m_a^2}{\phi}\int\frac{\dd^3p}{(2\pi)^3 2E_a}\dfa,
\end{equation}
where the field is coupled to the off-equilibrium particle distributions $\dfa$~\cite{Moore:1995si}. Here, $a$ corresponds to the particle degrees of freedom, one for $\Phi$ and four for $\Psi$. We emphasize that the description does not merely contain correct features but follows directly from the underlying QFT.

In the EoM for the classical field, Eq.~\eqref{eq:startingPointForFields}, the term on the right-hand side couples the field to the thermal bath of the off-equilibrium plasma particles. It provides stochastic kicks for the nucleating field, but also dissipation through backreaction, as we will see below in more detail.

Even though we refer to $\dfa$ as the off-equilibrium particle distributions, they are actually nonzero in any given realization at equilibrium, $\dfa\neq0$. This reflects equilibrium thermal fluctuations around the thermal mean, $\dfa=0$. In Sec.~\ref{sec:hamiltonian} below, we derive the equilibrium distribution for $\dfa$ to be Gaussian noise with the amplitude of an ideal quantum gas.%
\footnote{
In Secs.~\ref{sec:nucleationDistribution}, \ref{sec:offEquilDir}, we will derive the modifications to equilibrium fluctuations of $\dfa$ due to nucleation.
}
These fluctuations correctly introduce high-temperature stochastic kicks to the real-time evolution of the nucleating field.

For bosonic quantum fields, $\Phi$ in this case, it is important to stress the division of fluctuations in Eq.~\eqref{eq:startingPointForFields} above. Only the hard fluctuations with $E\sim T$ are correctly described as the particles in the bosonic distribution function. The equilibrium part of the particles enters the potential $\vnucl$ through dimensional reduction and the off-equilibrium part enters the off-equilibrium distribution function corresponding to $\Phi$. The infrared, $E\sim m_\Phi$, is described by the classical long-wavelength field $\phi$. Hence, the momentum integral on the right must be regulated from the infrared and the field on the left regulated from the ultraviolet in order to avoid double counting.

The thermal particles are described by particle distribution functions, $f_a$, whose time evolution is given by the collisionless Boltzmann equation~\cite{Arnold:1997gh,Blaizot:2001nr},%
\footnote{
It may be incautious to use the Boltzmann equation for the particles of the nucleating field due to negative-mass-squared (NMS) regions. However, NMS does not appear problematically in our analysis, as it is a small perturbation for $E\sim T$. Also, the source term in Eq.~\eqref{eq:linearizedVlasov} is consistent, as the equilibrium for $E\sim T$ is unscathed by NMS (see e.g.\ Refs.~\cite{Gould:2021ccf,Hirvonen:2022jba}).
}
\begin{equation}\label{eq:GeneralVlasovEquation}
    \left(\partial_t +\mathbf{v}\cdot\grad +\mathbf{F}_a\cdot\partial_{\mathbf{p}}\right)f_a=0,
\end{equation}
where $\mathbf{v}=\mathbf{p}/E_a$ is a velocity vector and 
\begin{equation}
    \mathbf{F}_a=-\grad E_a=-\frac{\grad m_a^2}{2E_a}
\end{equation}
is the force acting on particles.

In general, there would be a collision term on the right-hand side. However, the particle collisions are negligible for nucleation dynamics in this model. They only become important at longer length scales than the size of nucleating bubbles: $L\sim(y^4\ln y^{-1}T)^{-1}\gg m_\phi^{-1}$~\cite{Arnold:1997gh}.

In order to describe nucleation correctly, we need to be able to describe the thermal fluctuations correctly. Hence, the analysis of the following sections would not work with collision terms. This is due to the fact that they incorrectly introduce dissipation to the system without the corresponding fluctuations. Thus, fluctuations of the system would dampen out before nucleation may take place. We will still conjecture in Sec.~\ref{sec:conclusionanddiscussion} that our result is not limited to negligible collision terms.

We can split the distribution functions into equilibrium and off-equilibrium parts, $f_a=\feq+\dfa$, where the equilibrium part is given by
\begin{equation}
    \feq=\frac{1}{e^{\beta E_a}\mp_a 1}
\end{equation}
with the upper sign corresponding to bosons and lower to fermions. The equilibrium distribution is a function of the energy. As a consequence, it satisfies the following equation,
\begin{equation}
    \qty(\mathbf{v}\cdot\grad-\frac{\grad m_a^2}{2E_a}\cdot\partial_\mathbf{p})\feq=0,
\end{equation}
and is also time dependent through the field dependence of the masses, $E_a=\sqrt{\mathbf{p}^2+m_a^2(\phi)}$.

We can now find the time evolution for the off-equilibrium part appearing in the field EoM in Eq.~\eqref{eq:startingPointForFields},
\begin{equation}\label{eq:linearizedVlasov}
    \partial_t\dfa=-\qty(\mathbf{v}\cdot\grad-\frac{\grad m_a^2}{2E_a}\cdot\partial_\mathbf{p})\dfa-\frac{\feq'}{2E_a}\dv{m_a^2}{\phi}\Dot{\phi},
\end{equation}
where we have defined $\feq'=\partial_E \feq$. There is a source for the off-equilibrium particles, originating from the time dependence of the equilibrium distribution.

The force term coming from the gradient of the mass could be dropped for the example model given by Eqs.~\eqref{eq:initialLagrangian}--\eqref{eq:noMassHierarchies}. This is due to the term being suppressed by the light masses, $m_a\ll T$. In addition, the masses could be neglected from $\feq'$, as $m_a^2\ll \mathbf{p}^2\sim T^2$, with $\feq'$ becoming independent of the position and time. We will focus on this scenario more in Sec~\ref{sec:exponentialGrothUltrarelPlasma}. However, we will otherwise keep the masses for generality.

We now have the effective real-time description of our system capable of describing fluctuations on the length scale of nucleation containing classical fields and thermal, off-equilibrium particles, Eqs.~\eqref{eq:startingPointForFields},~\eqref{eq:linearizedVlasov}.

\section{Effective Hamiltonian and Equilibrium Distribution}\label{sec:hamiltonian}

Here, we will find the conserved Hamiltonian of the effective description. It also gives the equilibrium statistics for the system, describing the fluctuations of the classical field and off-equilibrium particle distributions.

We define a canonical momentum $\pi\equiv\Dot{\phi}$, and the equations of motion become
\begin{align}\label{eq:EquationsOfMotion}
\hspace{-0.4cm}
\begin{split}
    \Dot{\phi} &= \pi,\\
    \Dot{\pi} &= \grad^2 \phi - \vnucl'(\phi)-\sum_{a}\dv{m_a^2}{\phi}\int\frac{\dd^3p}{(2\pi)^3 2E_a}\dfa,\\
    \Dot{\dfa} &= -\qty(\mathbf{v}\cdot\grad-\frac{\grad m_a^2}{2E_a}\cdot\partial_\mathbf{p})\dfa-\frac{\feq'}{2E_a}\dv{m_a^2}{\phi}\pi,
\end{split}
\end{align}
where we have defined $\dot{{}}\equiv\partial_t$.

The Hamiltonian for the dynamics of the system is
\begin{align}
\begin{split}\label{eq:Hamiltonian}
    H&=\int\ddx\Bigg(\frac{1}{2}\pi^2+\frac{1}{2}(\grad\phi)^2+\vnucl(\phi)\\
    &\qquad\qquad\quad+\sum_{a}\frac{1}{2}\int\frac{\ddp}{(2\pi)^3}\frac{\dfa^2}{-\feq'}\Bigg).
\end{split}
\end{align}
Note that $\feq'<0$ so that the $\dfa^2$ term is positive definite. We present the Poisson brackets giving the time evolution from the Hamiltonian similarly to Refs.~\cite{Nair:1994xs,Iancu:1998bmf,Blaizot:2001nr} in Appendix~\ref{appendix:Hamiltonian}.

The conservation can be shown with a direct computation:
\begin{align}
\begin{split}
    &\quad\;\dv{H}{t}=\int\ddx\Big[\pi\qty(\Dot{\pi}-\grad^2\phi+\vnucl'(\phi))\\
    &\qquad\qquad\qquad\;\;\;\;+\sum_a\int\frac{\ddp}{(2\pi)^3}\frac{\dfa}{-\feq'}\Dot{\dfa}\Big]
\end{split}\\[5pt]
    &=\sum_a\int\frac{\ddx\ddp}{(2\pi)^3}\frac{\dfa}{\feq'}\qty(\mathbf{v}\cdot\grad-\frac{\grad m_a^2}{2E_a}\cdot\partial_\mathbf{p})\dfa\label{eq:afterManifestCancelationsHamCons}\\[5pt]
    &=\frac{1}{2}\sum_a\int\frac{\ddp}{(2\pi)^3}\mathbf{p}\cdot\underbrace{\int\ddx\,\grad\frac{\dfa^2}{E_a\feq'}}_{=0}\nonumber\\
    &\quad-\frac{1}{2}\sum_a\int\ddx\frac{\grad m_a^2}{2}\cdot\underbrace{\int\frac{\ddp}{(2\pi)^3}\,\partial_\mathbf{p}\frac{\dfa^2}{E_a\feq'}}_{=0}\\
    &\quad-\frac{1}{2}\sum_a\int\frac{\ddx\ddp}{(2\pi)^3}\dfa^2\underbrace{\qty(\mathbf{p}\cdot\grad-\frac{\grad m_a^2}{2}\cdot\partial_\mathbf{p})\frac{1}{E_a\feq'}}_{=0}\nonumber\\
&=0.\label{eq:HamiltonianConserved}
\end{align}
For the second equality, Eq.~\eqref{eq:afterManifestCancelationsHamCons}, we have used the equations of motion in Eq.~\eqref{eq:EquationsOfMotion}. After manifest cancellations, only the terms relating to the propagation of the off-equilibrium particles remain. We then used the product rule to split the terms into a sum of terms containing total derivatives and a manifestly zero term to show the total conservation of the Hamiltonian.

The equilibrium distribution is given by the Hamiltonian
\begin{equation}\label{eq:equilibriumDistribution}
    \rho_\text{eq}\propto e^{-\beta H},
\end{equation}
consisting of three uncorrelated parts for $\pi$, $\phi$ and $\dfa$.

The equilibrium distribution for $\phi$ was computed using the imaginary-time formalism in Ref.~\cite{Gould:2021ccf}, also containing $\vnucl$ as the potential. This establishes it as the correct potential for the $\phi$ field in Eq.~\eqref{eq:startingPointForFields}. Moreover, the link between the equilibrium distribution, Eq.~\eqref{eq:equilibriumDistribution}, and the EoM for $\phi$, Eq.~\eqref{eq:startingPointForFields}, demonstrates that the equilibrium potential can be used to study nucleation.

The particle fluctuations, $\dfa$, are spatially uncorrelated Gaussian noise in equilibrium.
Note that their equilibrium fluctuations match the free Bose and Fermi gases, $\langle\dfa^2\rangle=-T\feq'$, as they should.

\section{Critical Bubble and Exponential Growth}\label{sec:configurations}

There will be a large variety for nucleating configurations and their histories. In the Secs.~\ref{sec:nucleationDistribution}-\ref{sec:nucleationRate} below, we will account for all of them to obtain the nucleation rate. Here, we will look at two special solutions to the equations of motion: the critical bubble and the exponentially growing bubble. These solutions give lower bounds on the configurations that contribute to nucleation in terms of the Hamiltonian, Eq.~\eqref{eq:Hamiltonian}. As a consequence, their properties will appear in the final rate formula.

Discussing these solutions to the equations of motion gives important intuition regarding nucleation.  For this reason, it is also important to highlight that they are extremely uncharacteristic configurations: They are completely devoid of thermal fluctuations unlike practically all of the nucleating configurations.

\begin{figure}
    \centering
    \includegraphics[width=\columnwidth]{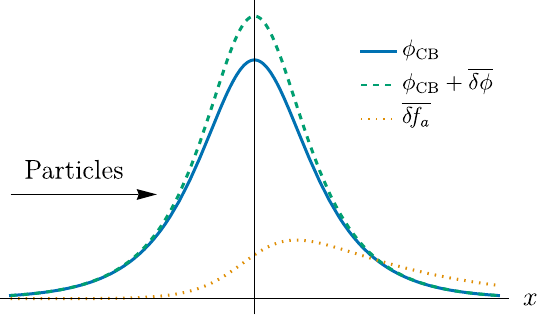}
    \caption{Critical bubble and the exponentially growing configuration. The shown off-equilibrium particles, $\ldf$, originate from the equilibrium particle flow from the left, which is pushed out of equilibrium by the exponentially growing bubble.}
    \label{fig:expGrowth}
\end{figure}

In a first-order phase transition, the equations of motion, Eq.~\eqref{eq:EquationsOfMotion}, permit for an important solution regarding nucleation, the critical bubble: $\phi=\cb$, $\pi=0$, $\dfa=0$. It is an unstable but static, spherically symmetric solution to
\begin{equation}
    \grad^2\cb=\vnucl'(\cb)
\end{equation}
interpolating between the phases, depicted in Fig.~\ref{fig:expGrowth}. It is the lowest-Hamiltonian configuration that is on the verge of nucleation. Hence, all of the nucleating configurations have a higher value of the Hamiltonian, resulting in the exponential suppression of nucleation.

We want to find the leading order nucleation rate. Hence, we expand in the field values around the critical bubble configuration, $\phi=\cb+\delta\phi$, to linear order.
Most notably, the EoM of $\pi$ becomes
\begin{align}\label{eq:phiEOMLinearizedAroundBubble}
    \Dot{\pi} = \qty(\grad^2 - \VCB'')\delta\phi-\sum_{a}\dv{m_a^2}{\phi}\int\frac{\dd^3p}{(2\pi)^3 2E_a}\dfa,
\end{align}
where we have defined $\VCB''\equiv\vnucl''(\phi_\text{CB})$.
Note that the squared masses are evaluated on the critical bubble in the linearized approximation, $m_a^2=m_a^2(\cb)$. Hence, their derivatives and the particle energies, $E_a$, are time independent. The same approximation for the masses also applies to the off-equilibrium-particle EoM in Eq.~\eqref{eq:EquationsOfMotion}, where also $\feq'$ is now evaluated on the critical bubble.

The Hamiltonian can be expanded around the critical bubble as well,
\begin{align}
    \hquad&=\HCB+\hfluct,\label{eq:CBHamiltonianQuad}\\
\begin{split}
    \HCB&=\int\ddx\qty(\frac{1}{2}(\grad\cb)^2+\vnucl(\cb)),\label{eq:CBHamiltonian}
\end{split}\\
\begin{split}
    \hfluct&=\int\ddx\Bigg(\frac{1}{2}\pi^2+\frac{1}{2}\delta\phi\qty(-\grad^2+\VCB'')\delta\phi\\
    &\qquad\qquad\quad+\sum_{a}\frac{1}{2}\int\frac{\ddp}{(2\pi)^3}\frac{\dfa^2}{-\feq'}\Bigg).\label{eq:HamiltonianFluct}
\end{split}
\end{align}
The linear term is zero due to expanding around the critical bubble. Fluctuations around the critical bubble are described by $\hfluct$.

The expanded Hamiltonian has one negative eigenmode,  $\delta\phi=\phi_-$, $\pi=0$, $\dfa=0$,
\begin{equation}\label{eq:negativeEigenmode}
    \qty(\grad^2 - \VCB'')\phi_-=\lambda_-\phi_-, \qquad \lambda_-<0.
\end{equation}
The Hamiltonian, $\hquad$, forms a barrier between the phases around the critical bubble. The negative-eigenmode direction in the phase space corresponds to the descent to the two phases from the critical bubble.

One can define similarly $\hmetaquad$, $\Hmeta$ and $\hmetafluct$ by expanding the Hamiltonian around the metastable phase, $\phi=\phi_\text{meta}$. The exponential suppression of nucleation comes from $\DHCB=\HCB-\Hmeta$, and $\hmetafluct$ describes fluctuations around the metastable phase.

The exponentially growing solution to the linearized equations, depicted in Fig.~\ref{fig:expGrowth}, will also be relevant for our analysis.
Since the system is linear, everything grows homogeneously with the same exponential factor:
\begin{equation}\label{eq:expGrowthTimeEvolution}
    \delta\phi=\ldp \,e^{\kappa t},\quad \pi=\lp\, e^{\kappa t},\quad \dfa=\ldf\, e^{\kappa t}.
\end{equation}

Here, we want to foreshadow the importance of the exponential growth rate, $\kappa$, in the final nucleation-rate formula of the full analysis in Eq.~\eqref{eq:finalResult}: It appears in the formula as a part of the prefactor and it will be the only part of the formula that is explicitly modified by the off-equilibrium particles.

The equations of motion for the exponentially growing solution become
\begin{align}
    \kappa^2\ldp&=\qty(\grad^2 - \VCB'') \ldp-\sum_{a}\dv{m_a^2}{\phi}\int\frac{\dd^3p}{(2\pi)^3 2E_a}\ldf,\label{eq:expGrowthPhi}\\
    \kappa\ldf&=-\qty(\mathbf{v}\cdot\grad-\frac{\grad m_a^2}{2E_a}\cdot\partial_\mathbf{p})\ldf-\frac{\feq'}{2E_a}\dv{m_a^2}{\phi}\kappa\ldp.\label{eq:expGrowthF}
\end{align}
From these equations, it is possible to solve for both $\kappa$ and the exponentially growing solutions, $\ldp$ and $\ldf$. Note that we already solved for $\lp=\kappa\ldp$.

The exponentially growing  bubble has a special property that the fluctuation part of the Hamiltonian, Eq.~\eqref{eq:HamiltonianFluct}, is zero, as we will show below. This will be an important piece of knowledge for showing that the exponential growth is slowed down by the thermal off-equilibrium particles and for deriving the rate in Sec.~\ref{sec:nucleationRate}. It also shows that the exponentially growing configuration has the same value for the Hamiltonian as the critical bubble and hence it can be regarded as the critically nucleated configuration.

The zero value follows from a rather brief derivation:
\begin{equation}\label{eq:zerHamiltonian}
\begin{split}
    \hfluct[\ldp,\lp,\ldf]&=\hfluct[\ldp e^{\kappa t},\lp e^{\kappa t},\ldf e^{\kappa t}]\\
    &=e^{2\kappa t}\hfluct[\ldp,\lp,\ldf]=0.    
\end{split}
\end{equation}
The first equality follows from the fact that the exponential evolution in Eq.~\eqref{eq:expGrowthTimeEvolution} satisfies the equations of motion, and hence the Hamiltonian is conserved. The second equality follows from the form of the Hamiltonian in Eq.~\eqref{eq:HamiltonianFluct}. Finally, the last equality follows from the independence of time and the positive value of $\kappa$: $e^{2\kappa t}\hfluct[\ldp,\lp,\ldf]$ can only be independent of time if $\hfluct[\ldp,\lp,\ldf]=0$.

Let us now understand $\kappa$ in more detail, and the off-equilibrium particle effects in particular.

In the absence of the off-equilibrium particles, $\dfa=0$, the exponentially growing configuration would match the negative eigenmode of the critical bubble, $\ldp=A\,\phi_-$, Eq.~\eqref{eq:negativeEigenmode}, with an arbitrary normalization, $A$. The exponential growth rate would be given by the negative eigenvalue, $\kappa=\sqrt{\abs{\lambda_-}}$.

We can prove that the off-equilibrium particles physically always slow down the exponential growth from $\sqrt{\abs{\lambda_-}}$ using the zero value of the fluctuation part of the Hamiltonian, Eq.~\eqref{eq:zerHamiltonian}.

Let us begin by choosing a specific normalization for $\ldp$, $\int\ddx\,\ldp^2=2A^2$, and expanding the exponentially growing configuration in the orthonormal eigenbasis of the fluctuation operator (\textit{cf.} Eq.~\eqref{eq:negativeEigenmode}),
\begin{equation}\label{eq:spectralDecompositionOfExpGrowth}
    \ldp=\sqrt{2}\,\sum_n a_n\phi_n.
\end{equation}
Then, we can rearrange the equation for the zero value of the Hamiltonian into the form of
\begin{align}\label{eq:growthProof}
\begin{split}
    A^2\kappa^2&=-\sum_n a_n^2\lambda_n -\sum_{a}\frac{1}{2}\int\frac{\ddx\ddp}{(2\pi)^3}\frac{\ldf^2}{-\feq'}\\
    &<a_-^2\abs{\lambda_-}<A^2\abs{\lambda_-}.
\end{split}
\end{align}
The first inequality follows from dropping the negative definite terms coming from the off-equilibrium particles and positive eigenmodes in the eigenexpansion. The final inequality can be proven by squaring and integrating over both sides of Eq.~\eqref{eq:spectralDecompositionOfExpGrowth}. From the two consecutive inequalities, we directly obtain $\kappa<\sqrt{\abs{\lambda_-}}$.

We can now identify two ways the off-equilibrium particles slow down the exponential growth: Firstly, they are sourced by the growth in the negative eigenmode, taking up some positive amount of the conserved Hamiltonian. In other words, the off-equilibrium particles backreact onto the bubble, slowing it down. Secondly, they then modify the growing field configuration away from the optimal negative eigenmode, hindering the growth even more.

\section{Exponential growth with ultrarelativistic plasma}\label{sec:exponentialGrothUltrarelPlasma}

Here, we will study the exponential growth even further. To make analytic progress, we will assume that the masses of the particles will be small in comparison to the temperature, $m_a\ll T$, so that all of the plasma particles are ultrarelativistic. Consequently, the force term in the off-equilibrium particle EoM, Eq.~\eqref{eq:linearizedVlasov}, can be neglected, as well as masses from the particle energies, $E=\abs{\mathbf{p}}\sim T$. We will find that generally the contributions from the light bosonic quantum fields are dominated by the corresponding classical infrared field, and that at least for the toy model given by Eqs.~\eqref{eq:initialLagrangian}--\eqref{eq:noMassHierarchies} the effects of the off-equilibrium particles constitute a perturbation. We also derive this leading correction to $\kappa$.

One can formally solve for $\ldf$ form Eq.~\eqref{eq:expGrowthF} without the force term, and insert it to Eq.~\eqref{eq:expGrowthPhi}. The ensuing energy integral can be computed in dimensional regularization to obtain
\begin{align}\label{eq:DimRegGrowth}
    \kappa^2\ldp&=\qty(\grad^2 - \VCB'') \ldp\\
    &\;+\sum_{a}\pm_a\dv{m_a^2}{\phi}\int\frac{\dd^2\Omega_\mathbf{v}}{(4\pi)^2 
    }\frac{\kappa}{\kappa+\mathbf{v}\cdot\grad}\qty[\dv{m_a^2}{\phi}\ldp].\nonumber
\end{align}
Here, $\int\dd^2\Omega_\mathbf{v}$ is the integral over the angle of the unit velocity $\mathbf{v}$, normalized to unity. The sign $\pm_a$ is positive for bosons and negative for fermions.

Note that here the bosonic contributions appear to make the exponential growth faster. This unphysical conclusion is a symptom of the infrared dominance of the light bosonic quantum fields. The leading nonlinear contributions from the corresponding classical field are bigger than those of the bosonic $E\sim T$ scale particles and hence the sign in dimensional regularization can differ from expected.%
\footnote{
One can perform the relevant energy integrals in cut-off regularization: $\int_{\Lambda_\text{IR}}^\infty\dd E f_{\text{eq,b}}'=-\frac{T}{\Lambda_\text{IR}}+\frac{1}{2}+\order{\Lambda_\text{IR}/T}$ and $\int_{\Lambda_\text{IR}}^\infty\dd E f_{\text{eq,f}}'=-\frac{1}{2}+\order{\Lambda_\text{IR}/T}$. Both are manifestly negative, corresponding to dissipation. For the former bosonic integral however, dimensional regularization only retains the finite positive part.
}
In the proof of the growth being inhibited in Eq.~\eqref{eq:growthProof}, the first inequality does not hold for bosons in dimensional regularization, because the $\ldf^2$ term turns out to be positive.

The same infrared dominance has been recently observed in Ref.~\cite{DeCurtis:2024hvh} for wall velocities. The nonlinear contributions from a classical field that complement the dimensionally regularized result above have been studied in a Langevin description in Ref.~\cite{Ekstedt:2022tqk}.

Finally, we will powercount the terms in the dimensionally regularized formula, Eq.~\eqref{eq:DimRegGrowth}, using the powercounting rules we laid out above in Eqs.~\eqref{eq:perturbativity}, \eqref{eq:noMassHierarchies}. We will note that the correction is perturbative, and find the leading correction to the exponential growth rate, $\kappa$.

Let us first estimate the change of the field value between the phases. Each of the potential terms in Eq.~\eqref{eq:nucleationPotential} are of the same order ($y^2T^4$) as long as $\Delta \phi\sim T$. Hence, this gives the powercounting estimate of the field-value change between the phases. The mass-squared derivatives can then be estimated from Eqs.~\eqref{eq:scalarMass}, \eqref{eq:fermionMass} as
\begin{equation}
    \dv{m_a^2}{\phi}\sim\frac{\Delta m_a^2}{\Delta \phi}\sim y^2T,
\end{equation}
which is the same for both the real scalar and the Dirac fermion.

Let us then look at the parameters related to the spatial and temporal characteristics of the bubble. The gradients can be estimated with the extent of the critical bubble, which can then be estimated with the mass of the nucleating field~\cite{Gould:2021ccf}:
\begin{equation}
    \grad\sim R_\text{CB}^{-1}\sim (\vnucl'')^{1/2}\sim y T.
\end{equation}
If we assume that the behavior of the critical bubble is not changed parametrically by the off-equilibrium particles, we obtain
\begin{gather}
    \kappa^2\sim\abs{\lambda_-}\sim\vnucl''\sim y^2T^2,\\
    \qty(\grad^2 - \VCB'') \sim \abs{\lambda_-}\sim y^2T^2.
\end{gather}

From the above estimates, we can gather that the first term in Eq.~\eqref{eq:DimRegGrowth} scales as $y^2T^2\ldp$ and is dominant to the off-equilibrium particle term scaling as $y^4T^2\ldp$. Thus, one can expect the equilibrium estimate of
\begin{equation}\label{eq:LeadingOrderApproximationToExpGrowth}
    \kappa= \sqrt{\abs{\lambda_-}} \times\qty(1+\order{y^2})
\end{equation}
to be valid.
This leading estimate is computed by \texttt{BubbleDet} when performing thermal nucleation rate computations~\cite{Ekstedt:2023sqc}.

We can then find the leading-order correction to the exponential growth by inserting $\kappa=\sqrt{\lambda_-}+\Delta\kappa$ and $\ldp=\phi_-+\Delta\ldp$ to the result in Eq.~\eqref{eq:DimRegGrowth}, and expanding to linear order in the perturbations, including the particle term. Integrating over both sides with the weight of $\phi_-$ yields
\begin{align}\label{eq:LeadingOrderCorrectionToGrowth}
    \Delta\kappa
    &=\sum_{a}\pm_a\frac{1}{2}\int\ddx\int\frac{\dd^2\Omega_\mathbf{v}}{(4\pi)^2 
    }\times\\
    &\qquad\qquad\quad\times\dv{m_a^2}{\phi}\phi_-\frac{1}{\sqrt{\abs{\lambda_-}}+\mathbf{v}\cdot\grad}\qty[\dv{m_a^2}{\phi}\phi_-].\nonumber
\end{align}
The linear $\Delta\ldp$ terms have canceled identically on both sides of the equation. The structure of the correction is that there is a source of the form $\dv{m_a^2}{\phi}\phi_-$, which creates a propagating disturbance in the particle distributions. This in turn affects the evolution of the bubble again through $\dv{m_a^2}{\phi}\phi_-$.

\section{Stationary Distribution for Nucleation}\label{sec:nucleationDistribution}

In order to take into account all of the possible configurations that nucleate, we introduce a probability distribution for the system, $\rho$. It obeys the Liouville equation,
\begin{align}\label{eq:MainLiouville}
\begin{split}
    \pdv{\rho}{t}&=-\int\ddx\Big(\Dot{\delta\phi}\funcdv{\delta\phi}+\Dot{\pi}\funcdv{\pi}\\
    &\qquad\qquad\quad\;\;+\sum_a\int\ddp\,\Dot{\dfa}\funcdv{\dfa}\Big)\rho,
\end{split}
\end{align}
where the time derivatives on the variables are given by the Hamiltonian equations in Eq.~\eqref{eq:EquationsOfMotion}.
(See Appendix~\ref{app:LiouvilleEquation}.)
Here, we will find the stationary solution to the Liouville equation that describes nucleation, whose boundary conditions are equilibrium in the metastable phase and zero in the stable phase.

The stationary distribution approximates suppressed nucleation. It is reached when the source for nucleation, i.e.\ the equilibrium distribution in the metastable state, remains approximately constant. The suppression is provided by the Boltzmann factor of the critical bubble (see Eq.~\eqref{eq:finalResult} below).

We will use Kramer's ansatz~\cite{Kramers:1940zz} also used by Refs.~\cite{Langer:1969bc,Berera:2019uyp,Ekstedt:2022tqk}. There, one assumes that the distribution is of the following form:
\begin{equation}
    \rho=\sigma(u)\times\frac{e^{-\beta \hquad}}{Z_\text{meta}}.
\end{equation}
The Boltzmann factor contains the equilibrium behavior around the critical bubble, and the function $\sigma$ encodes the off-equilibrium nature of the distribution. The partition function, $Z_\text{meta}$, is evaluated around the metastable phase because the system is in equilibrium there.

Kramer's ansatz also has an assertion that the out-of-equilibrium behavior, $\sigma(u)$, only depends on
\begin{align}\label{eq:definingu}
\begin{split}
    u&=\int\ddx\, U_{\delta\phi}\delta\phi+\int\ddx\, U_{\pi}\pi\\
    &\quad+\sum_a\int\ddx\ddp\, U_{\dfa}\dfa.
\end{split}
\end{align}
We show that this is a correct assertion by finding the coefficients $U$ in Sec.~\ref{sec:offEquilDir} in terms of the exponentially growing configuration.

The intuition for the assertion is that the system is only out of equilibrium in one direction in the phase space that interpolates between equilibrium, $\sigma=1$, in the metastable phase and no population, $\sigma=0$, in the stable phase. The function $u$ encodes this direction. In principle, $u$ could be a nonlinear function of the phase space. However, the system is linear. Consequently, $u$ is linear with the linear coefficients, $U$, determining the off-equilibrium direction.

The stationary Liouville equation for the ansatz distribution becomes
\begin{align}
    \Dot{u}\,\sigma'(u)=0,
\end{align}
where the time derivative on $u$ acts on $\delta\phi$, $\pi$ and $\dfa$.

Notice that the function we are solving for, $\sigma$, is only a function of $u$, but the coefficient in front is $\Dot{u}$. In order for Kramers' ansatz to hold, $\Dot{u}$ must be a function of $u$:
\begin{align}\label{eq:KramersCondition}
    \Dot{u}=\lambda u.
\end{align}
The linearity of the equality follows from the linearity of $u$ and $\Dot{u}$ in $\delta\phi$, $\pi$ and $\dfa$ This gives a condition for the off-equilibrium direction, $U$, that is solved for in Sec.~\ref{sec:offEquilDir}.

\begin{figure}
    \centering
    \includegraphics[width=\columnwidth]{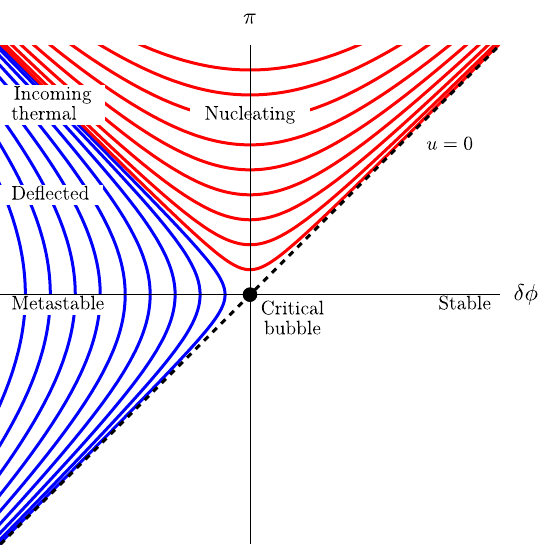}
    \caption{Two-dimensional slice of the phase space depicting the nucleating probability distribution around the critical bubble. Top-left corner corresponds to the $u<0$ region and to the equilibrium probability flow from the metastable phase. The curves depict phase-space trajectories, where the red curves correspond to trajectories that nucleate and the blue ones remain subcritical and are deflected back.}
    \label{fig:nucleatingDistribution}
\end{figure}

The Liouville equation simplifies to
\begin{align}
    \lambda u\,\sigma'(u)=0,
\end{align}
whose solution satisfying the relevant boundary conditions is the step function, $\sigma = \theta(-u)$. The full distribution, satisfying the boundary conditions for nucleation, is consequently
\begin{equation}\label{eq:NucleatingDistribution}
    \rho=\frac{\theta(-u)}{Z_\text{meta}}e^{-\beta \hquad}.
\end{equation}
The distribution is depicted in Fig.~\ref{fig:nucleatingDistribution}, and it contains an equilibrium flow toward the critical bubble configuration, $\cb$, from the metastable phase. Some of the configurations are able to pass over to the stable phase, i.e.\ nucleate, and some fail, being deflected back to the metastable phase.

\section{Off-Equilibrium Direction in the Phase Space}\label{sec:offEquilDir}

We will now solve for the $U$ coefficients in terms of the exponentially growing solution.

Equation~\eqref{eq:KramersCondition} has to hold for every value of $\delta\phi$, $\pi$ and $\dfa$. With the definition of $u$, Eq.~\eqref{eq:definingu}, we obtain three conditions that must hold:
\begin{align}
    \lambda U_{\delta \phi}&=\qty(\grad^2-\VCB'')U_\pi\label{eq:condForUDelPhi},\\
    \lambda U_\pi&=U_{\delta\phi}-\sum_a\int\ddp\frac{\feq'}{2E_a}\dv{m_a^2}{\phi} \,U_{\dfa},\\
\begin{split}
    \lambda U_{\dfa}&=\qty(\mathbf{v}\cdot\grad-\frac{\grad m_a^2}{2E_a}\cdot\partial_\mathbf{p}) U_{\dfa}\\
    &\quad-\frac{1}{(2\pi)^32E_a}\dv{m_a^2}{\phi} \,U_\pi.
\end{split}
\end{align}

We can solve for $U_{\delta \phi}$ from the middle equation, and insert the following ansatz for the remaining components of $U$ in terms of the exponentially growing configuration:
\begin{align}
    U_\pi=-\kappa\overline{\delta\phi},\quad U_{\dfa}=\frac{\overline{\dfa}}{(2\pi)^3\feq'},\quad \lambda=-\kappa.\label{eq:UsPiF}
\end{align}
This leads to the equations, which describe the exponentially growing configuration, Eqs.~\eqref{eq:expGrowthPhi}, \eqref{eq:expGrowthF}, thus, confirming the above ansatz.

We can now solve for $U_{\delta\phi}$ from Eq.~\eqref{eq:condForUDelPhi} and note the exponentially growing $\pi$ being $\lp=\kappa\ldp$,
\begin{align}
    \udp=\qty(\grad^2-\VCB'')\overline{\delta\phi},\quad U_\pi=-\kappa\ldp=-\lp.\label{eq:UsPhiPi}
\end{align}
We have now obtained the off-equilibrium direction, $U$, in Eqs.~\eqref{eq:UsPiF}, \eqref{eq:UsPhiPi}.

Notice that for example the exponential configuration lies on the $u=0$ surface. This follows directly from the form of $u$, Eq.~\eqref{eq:definingu}, \eqref{eq:UsPiF}, \eqref{eq:UsPhiPi} and the zero fluctuating Hamiltonian, Eq.~\eqref{eq:zerHamiltonian}.

\section{Nucleation Rate from Probability Current}\label{sec:nucleationRate}

The nucleation rate is the flux of a phase-space probability current, $\mathbf{J}$, over a phase-space surface between the two phases, $S$,
\begin{equation}
    \Gamma=\int \mathcal{D}\mathbf{S}\cdot \mathbf{J},
\end{equation}
where $\mathcal{D}\mathbf{S}$ is the phase-space area element orthogonal to the surface.
The probability current is given by the product of the phase-space velocity and the probability distribution,
\begin{equation}
    \mathbf{J} = 
    \begin{pmatrix}
    \Dot{\delta\phi} & 
    \Dot{\pi} & 
    \Dot{\delta f_a}
    \end{pmatrix}^\text{T} \rho,
\end{equation}
given in Eqs.~\eqref{eq:EquationsOfMotion}, \eqref{eq:phiEOMLinearizedAroundBubble}, \eqref{eq:NucleatingDistribution}.

We will define the surface, $S$, by
\begin{equation}\label{eq:definingSurface}
    \int\ddx\,\phi_-\,\delta\phi=0,
\end{equation}
i.e.\ by the field fluctuations around the critical bubble containing no negative eigenmode, Eq.~\eqref{eq:negativeEigenmode}, as in Refs.~\cite{Langer:1969bc,Gould:2021ccf,Hirvonen:2020jud}. Let us elaborate the choice.

For the nucleation rate, we want to obtain the contributions from all of the nucleating phase-space trajectories (see Fig.~\ref{fig:nucleatingDistribution} for intuition). Therefore, we must choose a surface through which they pass in order to nucleate. Similarly, the surface must be such that the non-nucleating trajectories begin and end on one side of the surface, giving no contribution to the rate. Note that the definition presented in Eq.~\eqref{eq:definingSurface} is not unique in this regard, but it will be easy to implement computationally and is physically intuitive.

There is a barrier between the metastable and stable phases in the phase space, which is given by the Hamiltonian, $\hquad$ in Eq.~\eqref{eq:CBHamiltonianQuad}, and located around the critical bubble, $\cb$. The barrier is concave in one direction in the phase space given by the negative eigenmode, which corresponds to the descent to the phases from the barrier. Since the concave direction is given by the negative eigenmode, the top of the barrier between the phases corresponds to the chosen surface. The nucleating phase-space trajectories must start on the metastable side of the barrier and end on the stable side while nucleating, hence passing through the surface. Correspondingly, the non-nucleating trajectories must begin and end on the metastable side of the surface.

The integral over the surface can now be made more explicit:
\begin{align}
    \Gamma&=\int \mathcal{D}S
    \begin{pmatrix}
    \phi_- & 
    0 & 
    0
    \end{pmatrix}\cdot \mathbf{J}\\
    &=\int \mathcal{D}S \frac{\theta(-u)}{Z_\text{meta}}e^{-\beta \hquad} \int\ddx\, \phi_-\,\pi.
\end{align}
Here, $\begin{pmatrix}
\phi_- & 
0 & 
0
\end{pmatrix}$ is the unit vector normal to the surface and $\phi_-$ is the normalized negative eigenmode.

A direct integration would be cumbersome because the step function, $\theta(-u)$, does not lead to a neat integration boundary, and there is a factor of $\pi$ in the integrand.

Physically, the convoluted boundary is related to the fact that particle fluctuations can make a configuration pass through the surface multiple times.%
\footnote{
Each nucleating phase-space trajectory is still counted only once to the rate in the integration. The trajectories carry thermal weight. Therefore, passing back to the metastable side cancels exactly previously entering the stable side. Due to this, the non-nucleating configurations give a zero contribution to the rate overall.
}
In the absence of off-equilibrium particles, the integration boundary on the surface $S$ would simply be $\int\ddx\, \phi_-\,\pi>0$, corresponding to the surface being crossed only once~\cite{Hirvonen:2020jud}. This would also make the direct evaluation of the integral straightforward. Note that the depiction in Fig.~\ref{fig:nucleatingDistribution} is in the $\dfa=0$ plane. Thus, the fluctuations do not cause multiple crossings for the depicted trajectories.

The strategy we adopt for the integral is to add an auxiliary term to the integrand:
\begin{equation}
    \tilde{\mathbf{J}}=\mathbf{J}+\begin{pmatrix}
1 & 
0 & 
0
\end{pmatrix}^\text{T}T\funcdv{\pi}\rho.
\end{equation}
The additional term is a total derivative of $\pi$, and vanishes from the integral as a boundary term because $\pi$ is integrated over on the surface $S$:
\begin{equation}
    \int \mathcal{D}\mathbf{S}\cdot \tilde{\mathbf{J}}=\int \mathcal{D}\mathbf{S}\cdot \mathbf{J}=\Gamma.
\end{equation}
The rate simplifies to having a delta function:
\begin{align}\label{eq:rateFinalIntegralContinuum}
    \Gamma= \frac{\kappa T\qty(\int\ddx\, \phi_-\,\ldp)}{Z_\text{meta}}\int \mathcal{D}S\delta(u)e^{-\beta \hquad} .
\end{align}

In the rest the section, we present the steps for the integration. The resulting nucleation rate is shown in Eq.~\eqref{eq:finalResult}.

For the integration, it is convenient to regulate the system by putting it on a finite lattice with $N$ spatial sites and $N$ momentum sites. The fine details are unimportant as this is just a convenient tool for computation, and its effects will disappear from the final result expressed in terms of the initial system.

The field and the canonical momentum can be expanded in the eigenbasis of the field dependent Hamiltonian (similarly to the negative eigenmode in Eq.~\eqref{eq:negativeEigenmode}) and the deviations of the particle distributions on the lattice sites,
\begin{align}\label{eq:discretizingVariables}
    \delta \phi &= \sum_{i=1}^N \xi_i \phi_i,\quad
    \pi = \sum_{i=1}^N \xi_{N+i} \phi_i,\quad
    \dfa=\sum_{i=1}^{N^2}\xi_{2N+i}\delta_i,
\end{align}
where $\phi_i$ are the normalized eigenmodes, and the functions $\delta_i$ are one on their corresponding lattice sites and zero otherwise. We have now gathered everything into a single vector $\xi$, whose first element corresponds to the negative eigenmode.

The fluctuation part of the Hamiltonian and the $u$ function from Eqs.~\eqref{eq:HamiltonianFluct}, \eqref{eq:definingu} become
\begin{equation}\label{eq:discretizedHamiltonianAndu}
    \hfluct=\frac{1}{2}\sum_i \lambda_i\xi_i^2,\quad u=-\sum_i\lambda_i\overline{\xi}_i\xi_i,
\end{equation}
where $\lambda_i$ are the eigenvalues of the discretized Hamiltonian, and $\overline{\xi}_i$ are the $\xi_i$ coefficient in Eq.~\eqref{eq:discretizingVariables} when setting $\delta\phi=\ldp$, $\pi=\lp$ and $\dfa=\ldf$.

In these discretized coordinates, the surface $S$ corresponds to $\xi_1=0$, and the rate in Eq.~\eqref{eq:rateFinalIntegralContinuum} becomes:
\begin{align}
    \Gamma=\frac{\kappa T\overline{\xi}_1 e^{-\beta \DHCB}}{\int\qty(\prod_i\dd\xi_i^\text{meta})e^{-\beta \hmetafluct}}\int\prod_{i>1}\dd\xi_i\,\delta(u)e^{-\beta \hfluct}.
\end{align}
The denominator follows from expanding the partition function $Z_\text{meta}$ around the metastable phase.

The denominator can be computed straightforwardly:
\begin{equation}
    \int\prod_i\dd\xi_i^\text{meta}e^{-\beta \hmetafluct}=\qty(\prod_i\frac{\lambda_i^\text{meta}}{2\pi T})^{-1/2}.
\end{equation}
Notice that only the eigenvalues corresponding to the field $\delta\phi$ are actually different from the eigenvalues $\lambda_i$.

The remaining nontrivial part is the integral
\begin{equation}
    I=\int\prod_{i>1}\dd\xi_i\,\delta(u)e^{-\beta \hfluct}.
\end{equation}

There are three zero modes, $\xi_2$, $\xi_3$, and $\xi_4$, related to translations. Note that neither $u$ nor $H$ depends on these. $H$ is translationally invariant and $u$ only depends on the spherically symmetric field eigenmodes. Thus, their contribution is just the volume of the zero-mode subspace $\mathcal{V}=V \DHCB^{3/2}$ that can be obtained using collective coordinates~\cite{GERVALS1976281,Vainshtein:1981wh}, where $V$ is the spatial volume of the system. The result is
\begin{equation}
    I=\mathcal{V}\int\prod_{i\geq5}\dd\xi_i\,\delta(u)e^{-\beta \hfluct}.
\end{equation}

We can regulate the delta function to obtain
\begin{align}
    \frac{I}{\mathcal{V}}&=\lim_{\epsilon\to0^+}\frac{1}{\epsilon\sqrt{2\pi}}\int\prod_{i\geq5}\dd\xi_i\,\exp(-\frac{u^2}{2\epsilon^2}-\beta \hfluct)\\
    &=\lim_{\epsilon\to0^+}\frac{1}{\epsilon\sqrt{2\pi}}\int\prod_{i\geq5}\dd\xi_i\,\exp\Big(-\frac{1}{2}\sum_{ij}Q_{ij}\xi_i\xi_j\Big)\\
    &=\lim_{\epsilon\to0^+}\frac{1}{\epsilon\sqrt{2\pi}}{\det}''\qty(\frac{Q}{2\pi})^{-1/2},
\end{align}
where the double prime refers to dropping the eigenvalues $i\in\{1,2,3,4\}$, and where we have defined the matrix
\begin{align}
    Q_{ij} &=\beta\lambda_i\delta_{ij}+\frac{\lambda_i\lambda_j\overline{\xi}_i\overline{\xi}_j}{2\epsilon^2}\\
    &={\sum_{nm}}^{\prime\prime}\sqrt{\beta\lambda_i}\delta_{in}\qty(\delta_{nm} +\frac{\sqrt{\lambda_n\lambda_m}\,\overline{\xi}_n\overline{\xi}_m}{2\beta\epsilon^2}) \sqrt{\beta\lambda_j}\delta_{mj}.\nonumber
\end{align}

Using the latter representation for $Q$, we can see that the determinant factorizes into two parts. There will be a product over the eigenvalues and the determinant of the Hermitian matrix, $\delta_{ij} +\frac{\sqrt{\lambda_i\lambda_j}\,\overline{\xi}_i\overline{\xi}_j}{2\beta\epsilon^2}$, whose eigenvalues are ones, apart from the eigenvalue of $1+\sum_{i\geq5}\frac{\lambda_i\overline{\xi}_i^2}{2\beta\epsilon^2}$ corresponding to the eigenvector of $\sqrt{\lambda_i}\,\overline{\xi}_i$. Thus, the integral becomes
\begin{align}\label{eq:withInfiniteSum}
    \frac{I}{\mathcal{V}}&=\frac{1}{\sqrt{2\pi T}}\qty(\sum_{i\geq5}\frac{\lambda_i\overline{\xi}_i^2}{2})^{-1/2}\qty(\prod_{i\geq5}\frac{\lambda_i}{2\pi T})^{-1/2},
\end{align}
where we have been able to take the limit $\epsilon\to0^+$.

Now the zero fluctuation Hamiltonian of the exponentially growing solution, Eq.~\eqref{eq:zerHamiltonian}, comes in handy. The sum in Eq.~\eqref{eq:withInfiniteSum} can be reformulated in terms of it,
\begin{align}
    \sum_{i\geq5}\frac{\lambda_i\overline{\xi}_i^2}{2}&=\hfluct[\ldp,\lp,\ldf]-\frac{\lambda_1\overline{\xi}_1^2}{2}
    =\frac{\abs{\lambda_-}\overline{\xi}_1^2}{2},
\end{align}
where the first equality follows from the discretized Hamiltonian in Eq.~\eqref{eq:discretizedHamiltonianAndu}, and where we have denoted the negative eigenvalue with $\lambda_-$.

The full rate formula becomes
\begin{align}
    \Gamma=\frac{\kappa T^{1/2}}{\sqrt{2\pi\abs{\lambda_-}}}\mathcal{V}\frac{\prod_{i\geq5}\qty(\frac{\lambda_i}{2\pi T})^{-1/2}}{\prod_i\qty(\frac{\lambda_i^\text{meta}}{2\pi T})^{-1/2}}e^{-\beta \DHCB}.
\end{align}
Remembering that the only differing eigenvalues are related to the field, we obtain
\begin{align}\label{eq:finalResult}
    \Gamma=\frac{\kappa}{2\pi}\frac{\mathcal{V}}{(2\pi T)^{\frac{3}{2}}}\abs{\frac{\det(-\grad^2+V_\text{meta}'')}{{\det}'\qty(-\grad^2+\VCB'')}}^{\frac{1}{2}}e^{-\beta \DHCB},
\end{align}
where the single prime refers to omitting the zero modes.

The result coincides with Langer's formula with the QFT-specific exponential growth rate, $\kappa$, of the exponentially growing configuration, Eqs.~\eqref{eq:expGrowthPhi}, \eqref{eq:expGrowthF}. When accounting for all nucleation histories, the off-equilibrium particle contributions average out, leaving modifications only to this growth rate. Notably, the equilibrium statistical part, computed with the EFT approach in Ref.~\cite{Gould:2021ccf}, factorizes from the rate even in this real-time computation and still governs the leading-order terms in the logarithm of the rate.

\section{Conclusion and Discussion}\label{sec:conclusionanddiscussion}

We have derived the nucleation rate formula for high-temperature QFTs, Eq.~\eqref{eq:finalResult}, in a direct real-time approach. In particular, we have taken into account the stochastic kicks and dissipation effects of out-of-equilibrium plasma during nucleation.

In the final result, the aforementioned effects only modify the exponential growth rate, $\kappa$, Eqs.~\eqref{eq:expGrowthPhi}, \eqref{eq:expGrowthF}. Crucially, they do not affect the nucleation rate exponentially, unlike the critical bubble, and possibly the fluctuation determinants~\cite{Ekstedt:2023sqc}.

Apart from the QFT-specific exponential growth rate, the resulting formula agrees with Langer~\cite{Langer:1969bc} and the EFT approach to thermal nucleation~\cite{Gould:2021ccf}.

The result applies directly to models, where the kinetic description for the particles is given by a collisionless Boltzmann equation, Eq.~\eqref{eq:GeneralVlasovEquation}.
The model dependency is in the exponentially growing configuration, Eqs.~\eqref{eq:expGrowthPhi}, \eqref{eq:expGrowthF}, having different particle content, and possibly other classical light bosonic fields. However, one should bear in mind that the dominant effects for the damping from light bosonic fields come from the infrared fields, not from the hard particles, as discussed in Sec.~\ref{sec:exponentialGrothUltrarelPlasma}.

In addition, we performed the powercounting in the model defined in Sec.~\ref{sec:ModelAndVlasov}. There, the leading order $\kappa$ is given by the negative eigenmode, Eq.~\eqref{eq:LeadingOrderApproximationToExpGrowth}, with the off-equilibrium particles giving a correction shown in Eq.~\eqref{eq:LeadingOrderCorrectionToGrowth}. The leading estimate can be numerically computed with \texttt{BubbleDet}~\cite{Ekstedt:2023sqc}.

The derivation here is restricted to the collisionless Boltzmann equation, because it is a nondissipative description, and we can construct a conserved Hamiltonian. Due to the universality of Langer's formula, we conjecture that the resulting rate formula also holds with the (linearized) collision term, which would then appear in the formula for the exponentially growing configuration in Eq.~\eqref{eq:expGrowthF}. This is supported by the fact that fluctuations around the configuration do not play a role in its growth rate, $\kappa$. It only includes the dissipation through off-equilibrium particles, which is correctly described by the Boltzmann equation with collisions. Also note that the underlying thermal QFT still satisfies the fluctuation-dissipation relation, even though the Boltzmann equation only captures correctly the dissipative aspects. This conjecture would increase the applicability to more realistic models with significant collision terms. In addition, the significant collision terms could alter the powercounting performed for the toy model and change parametrically the resulting rate.

\begin{acknowledgments}
We would like to thank
A.~Ekstedt,
O.~Gould,
A.~Guiggiani,
M.~Hindmarsh,
B.~Laurent,
L.~Niemi,
K.~Rummukainen,
P.~Schicho,
J.~van de Vis,
and A.~Vuorinen
for enlightening
discussions and valuable comments on the manuscript.
The work has been supported in part by the Academy of Finland Grant No.~354533 and by a Royal Society Dorothy Hodgkin Fellowship.
\end{acknowledgments}

\appendix

\section{Hamiltonian Description} \label{appendix:Hamiltonian}

\newpage
The nonzero canonical Poisson-bracket relations for the Hamiltonian description are
\begin{align}
    \{\phi(\mathbf{x}),\pi(\mathbf{y})\}&=\delta(\mathbf{x}-\mathbf{y}),\\
    \{\pi(\mathbf{x}),\dfa(\mathbf{y},\mathbf{q})\}&=\frac{\feq'}{2E_a}\dv{m_a^2}{\phi}\delta(\mathbf{x}-\mathbf{y}),\\
    \{\dfa(\mathbf{x},\mathbf{p}),\dfa(\mathbf{y},\mathbf{q})\}&=-(2\pi)^3\feq'\nonumber\\
    &\qquad\times\qty(\mathbf{v}\cdot\partial_\mathbf{x}+\mathbf{F}_a\cdot\partial_\mathbf{p})\\
    &\qquad\times\delta(\mathbf{p}-\mathbf{q})\delta(\mathbf{x}-\mathbf{y}).\nonumber
\end{align}
The time evolution, also given in Eq.~\eqref{eq:EquationsOfMotion}, can then be obtained from
\begin{align}
    \Dot{\phi}&=\{\phi,H\},\\
    \Dot{\pi}&=\{\pi,H\},\\
    \Dot{\dfa}&=\{\dfa,H\}.
\end{align}

\section{Liouville Equation}\label{app:LiouvilleEquation}

A phase-space probability distribution for the system obeys a Liouville equation, Eq.~\eqref{eq:MainLiouville}, if the phase-space probability flow is incompressible. The incompressibility can be shown with a direct computation,
\begin{align}
    &\quad\int\ddx\qty(\funcdv{\delta\phi}\Dot{\delta\phi}+\funcdv{\pi}\Dot{\pi}+\sum_a\int\ddp\funcdv{\dfa}\Dot{\dfa})\nonumber\\
    &=-\sum_a\int\ddx\int\ddp\Big(\mathbf{v}\cdot\underbrace{\funcdv{\dfa(\mathbf{x},\mathbf{p})}\grad\dfa(\mathbf{x},\mathbf{p})}_{=0},\nonumber\\
    &\qquad\qquad\qquad\qquad\qquad+\mathbf{F}_a\cdot\underbrace{\funcdv{\dfa(\mathbf{x},\mathbf{p})}\partial_\mathbf{p}\dfa(\mathbf{x},\mathbf{p})}_{=0}\Big)\nonumber\\
    &=0,
\end{align}
where we have used the equations of motion in Eq.~\eqref{eq:EquationsOfMotion}. The final step follows from $\delta'(0)=0$: Changing the function by a delta function at a point does not change the derivative at the point.

\input{main.bbl}


\end{document}

%% file: main.bbl
%